\documentclass[twocolumn,aps,prl,superscriptaddress]{revtex4}
\usepackage{graphicx,amsmath}
\newcommand{\ket}[1]{| #1 \rangle}
\newcommand{\bra}[1]{\langle #1 |}

\begin{document}
\title{Using Sideband Transitions for Two-Qubit Operations\\
in Superconducting Circuits}
\author{P.~J.~Leek}
\affiliation{Department of Physics, ETH Zurich, CH-8093, Z\"urich,
Switzerland.}
\author{S.~Filipp}
\affiliation{Department of Physics, ETH Zurich, CH-8093, Z\"urich,
Switzerland.}
\author{P.~Maurer}
\affiliation{Department of Physics, ETH Zurich, CH-8093, Z\"urich,
Switzerland.}
\author{M.~Baur}
\affiliation{Department of Physics, ETH Zurich, CH-8093, Z\"urich,
Switzerland.}
\author{R.~Bianchetti}
\affiliation{Department of Physics, ETH Zurich, CH-8093, Z\"urich,
Switzerland.}
\author{J.~M.~Fink}
\affiliation{Department of Physics, ETH Zurich, CH-8093, Z\"urich,
Switzerland.}
\author{M.~G\"oppl}
\affiliation{Department of Physics, ETH Zurich, CH-8093, Z\"urich,
Switzerland.}
\author{L.~Steffen}
\affiliation{Department of Physics, ETH Zurich, CH-8093, Z\"urich,
Switzerland.}
\author{A.~Wallraff}
\affiliation{Department of Physics, ETH Zurich, CH-8093, Z\"urich,
Switzerland.}
\date{\today}
\begin{abstract}
We demonstrate time resolved driving of two-photon blue sideband
transitions between superconducting qubits and a transmission line
resonator. Using the sidebands, we implement a pulse sequence that
first entangles one qubit with the resonator, and subsequently
distributes the entanglement between two qubits. We show generation
of 75\% fidelity Bell states by this method. The full density matrix
of the two qubit system is extracted using joint measurement and
quantum state tomography, and shows close agreement with numerical
simulation. The scheme is potentially extendable to a scalable
universal gate for quantum computation.
\end{abstract}
\maketitle

In the pursuit of a scalable architecture for quantum information
processing, Josephson-junction based superconducting circuits are
currently a leading candidate \cite{Clarke2008}. Controlled coherent
two-qubit interactions, essential for the realization of universal
quantum computation, have been demonstrated using a variety of
coupling schemes
\cite{Yamamoto2003,Steffen2006,Hime2006,Plantenberg2007,Niskanen2007,Sillanpaa2007,Majer2007}.
A promising recent advance in the field has been the realization of
circuit quantum eletrodynamics (QED) \cite{Wallraff2004} in which
superconducting qubits are strongly coupled to single photons in a
transmission line resonator. This architecture has been used to
couple two qubits over a distance of several millimeters by exchange
of a virtual photon between two resonant transmon qubits detuned
from the resonator \cite{Majer2007} and by exchange of a real photon
between two phase qubits in resonance with the resonator
\cite{Sillanpaa2007}.
The possibility to achieve such non-local couplings is advantageous
for the realization of multi-qubit systems.
It is also desirable to be able to switch two-qubit couplings on and
off with high accuracy, such that high fidelity operations can be
carried out both on individual qubits, and on specifically chosen
pairs of qubits, avoiding spurious couplings to other qubits in the
system. This issue has been addressed in superconducting circuits by
controlling the interaction strength between two qubits via an
external parameter \cite{Hime2006,Niskanen2007,Ploeg2007}, or by
tuning transitions into and out of resonance
\cite{Steffen2006,Sillanpaa2007,Majer2007}.

In this Letter we present an experimental realization of a two-qubit
coupling scheme in circuit QED that makes use of sideband
transitions between the joint qubit and resonator states. In this
scheme, qubit transition frequencies may be kept at values chosen
for optimal coherence, and do not need to be frequency tuned. Since
the coupling is only present when strong microwave drives are turned
on to induce the sideband transitions the tunability or on/off ratio
of the coupling can be large. The scheme is hence potentially
scalable in a similar manner to that achieved with sidebands in
trapped ion systems \cite{Schmidt-Kaler2003,Haffner2005}.

\begin{figure}[b]
\includegraphics[width=1.0 \columnwidth]{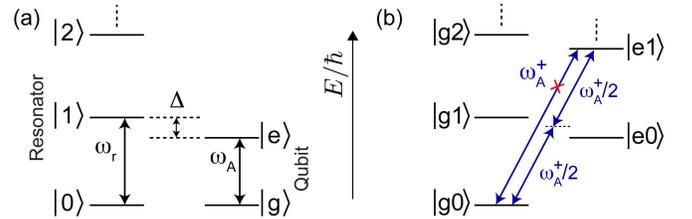}
\caption{(a) Schematic of the energy levels of a harmonic oscillator
with resonant frequency $\omega_{\rm r}$, and levels $\ket{n}$
corresponding to photon number $n$, and qubit with transition
frequency $\omega_{\rm A}$, and ground and first excited states
$\ket{g}$ and $\ket{e}$ respectively. (b) Combined level diagram,
indicating the blue sideband transition $\omega_{\rm
A}^+=\omega_{\rm r}+\omega_{\rm A}$, which is forbidden to first
order, but may be driven using two photons, for example both at a
frequency of $\omega_{\rm A}^+/2$.} \label{Fig:Sidebands}
\end{figure}

In the dispersive regime of circuit QED, the qubit transition
frequency $\omega_{\rm A}$ is detuned from the resonator frequency
$\omega_{\rm r}$ by $\Delta=|\omega_{\rm r}-\omega_{\rm A}|\gg g$,
where $g$ is the coupling strength between resonator and qubit on
resonance (see Fig.~\ref{Fig:Sidebands}(a)). Although the qubit and
resonator do not directly exchange energy in this case, the residual
dispersive coupling still allows sideband transitions linking the
qubit and resonator states to be accessed using strong additional
microwave drive fields \cite{Wallraff2007,Blais2007}. Blue sideband
transitions involve the simultaneous excitation of both qubit and
resonator, at a transition frequency $\omega_{\rm A}^+=\omega_{\rm
r}+\omega_{\rm A}$, while the red sideband involves the exchange of
an excitation between the two systems, at a transition frequency
$\omega_{\rm A}^-=|\omega_{\rm r}-\omega_{\rm A}|$. Such sideband
transitions were first observed in a superconducting circuit in an
experiment coupling a flux qubit to a superconducting quantum
interference device (SQUID) \cite{Chiorescu2004}.

Due to symmetry considerations in the circuit QED Hamiltonian,
single photon sideband transitions with either a Cooper Pair Box
(CPB) \cite{Bouchiat1998} biased at charge degeneracy, or with a
transmon qubit \cite{Koch2007,Schreier2008}, are forbidden to first
order. They may however be accessed using two photons, whose sum or
difference frequency is equal to one of the sideband transition
frequencies $\omega_{\rm A}^{+/-}$ \cite{Blais2007}. A spectroscopic
investigation of these two-photon transitions for a CPB has been
reported previously, in which two microwave drive fields of
different frequency were applied to the resonator, detuned slightly
from the qubit and resonator frequencies \cite{Wallraff2007}. The
increased coupling strength of the transmon to the resonator
\cite{Koch2007} when compared to the CPB allows driving of the
sidebands at larger detunings of the drives from the qubit and
resonator. Here we choose to drive the blue sideband with two
photons of equal energy, using a single microwave drive of frequency
$\omega_{\rm A}^+/2=(\omega_{\rm r}+\omega_{\rm A})/2$ (see
Fig.~\ref{Fig:Sidebands}(b)). In this configuration the drive is
equally detuned from the qubit and resonator, maximizing the
selectivity of the sideband transition with respect to undesired
off-resonant driving of the bare qubit transition and off-resonant
population of the resonator.

\begin{figure}[t]
\includegraphics[width=0.95 \columnwidth]{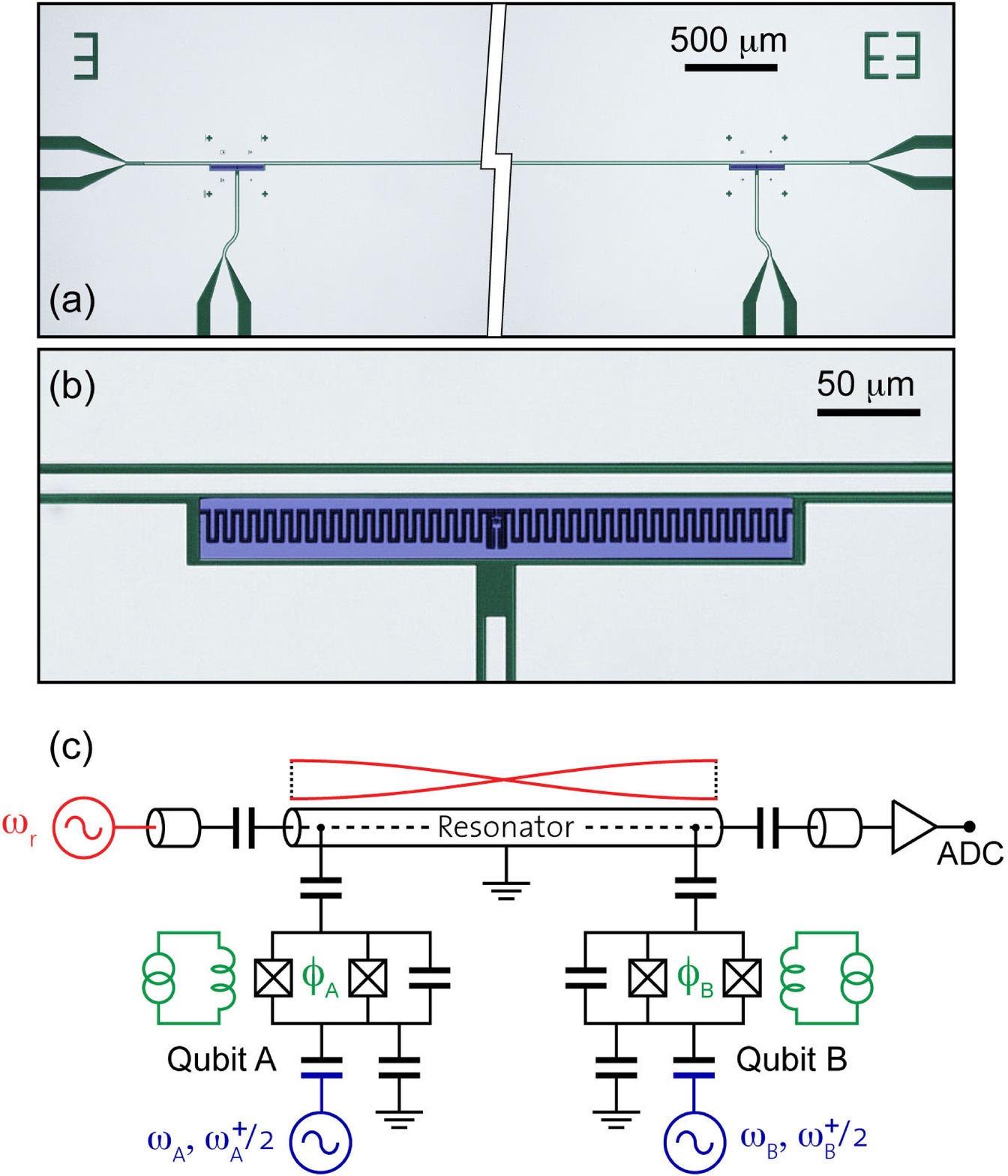}
\caption{(a) False color optical microscope image of the two-qubit
circuit QED device. The sapphire substrate is shown in dark green,
the Niobium resonator in light blue, and the Aluminium qubits in
dark blue. (b) High magnification false color optical microscope
image of one of the qubits coupled to the transmission line
resonator, and a local microwave drive line. (c) Electrical circuit
representation of the two-qubit circuit QED device with local
controls. The magnetic fluxes $\Phi_{\rm A}$ and $\Phi_{\rm B}$
through the SQUID loops of the qubits are tuned using local
superconducting coils (green), while transitions are driven using
local microwave drive lines (blue).} \label{Fig:Device}
\end{figure}

We work with a two qubit circuit QED device, for which an optical
image and circuit schematic are depicted in Fig.~\ref{Fig:Device}. A
transmon qubit is fabricated at each end of a coplanar waveguide
resonator with bare fundamental resonant frequency $\omega_{\rm
r}/2\pi=6.44~\rm{GHz}$. The qubits, labelled hereafter as qubit A
and B are near identical, with a coupling strength to the resonator
$g/2\pi=133~\rm{MHz}$ and a charging energy $E_{\rm
C}/h=232~\rm{MHz}$. The transition frequencies of the qubits are
tuned independently by changing the local magnetic fluxes $\Phi_{\rm
A}$ and $\Phi_{\rm B}$ through the SQUID loops of qubits A and B
respectively, using two small superconducting coils mounted below
the chip. Direct qubit and sideband transitions are driven
selectively using microwave transmission lines coupled capacitively
to each qubit locally (see Fig.~\ref{Fig:Device}(c)). For the
experiment, the qubits are tuned to transition frequencies of
$\omega_{\rm A}/2\pi=4.50~\rm{GHz}$ and $\omega_{\rm
B}/2\pi=4.85~\rm{GHz}$ respectively, well into the dispersive
regime.

\begin{figure}[b]
\includegraphics[width=1.0 \columnwidth]{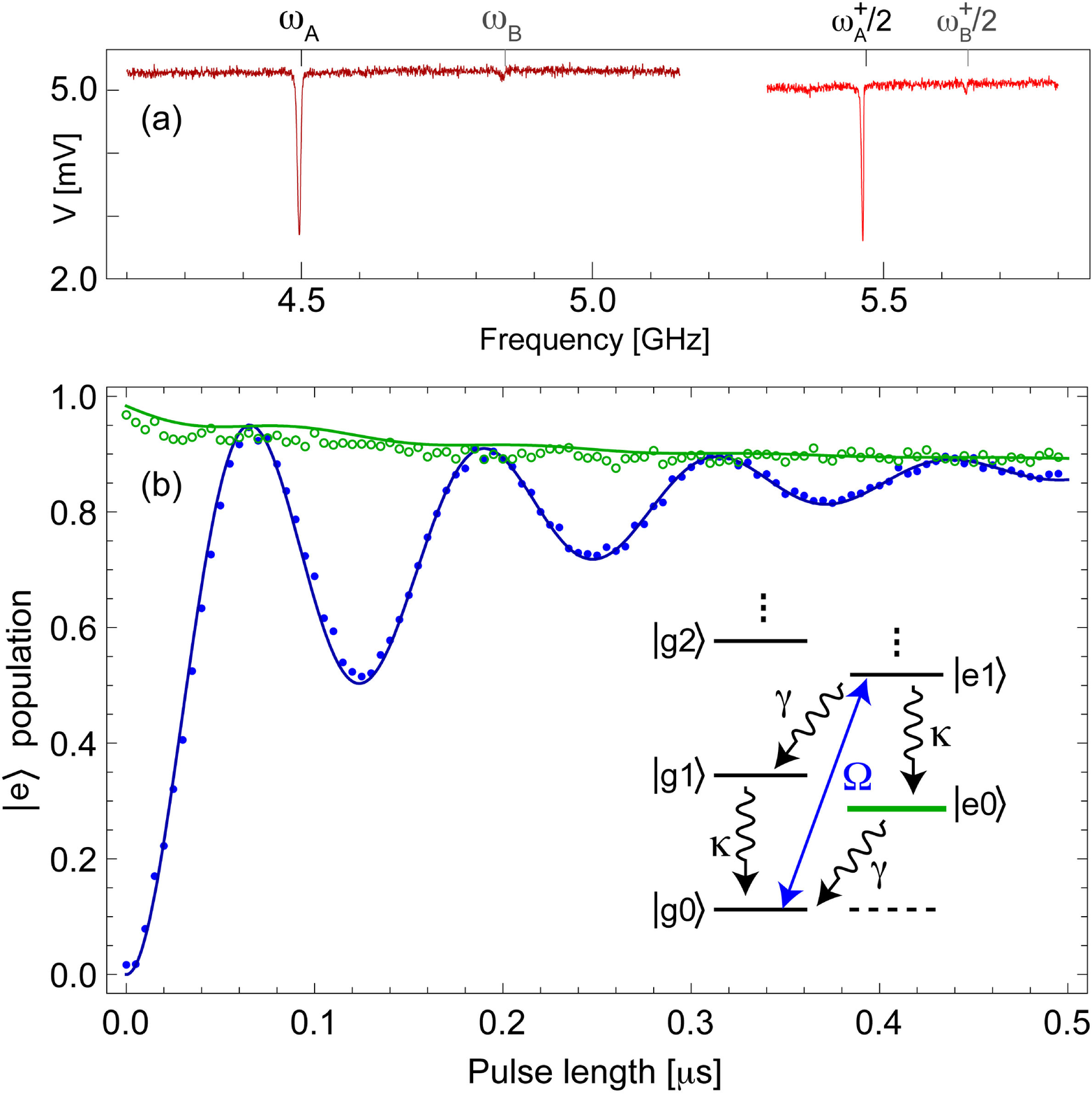}
\caption{(a) Spectroscopy on the selective drive line of qubit A,
showing the direct transition of the qubit at 4.50 GHz, and the
two-photon blue sideband at 5.47 GHz. (b) Time resolved measurements
of the excited state population of qubit A as a function of the
length of a blue sideband pulse. Blue solid circles show the
response for the initial state $\ket{g0}$, and green open circles
for $\ket{e0}$. Solid lines: Master equation simulations using decay
parameters extracted from separate measurements.
(\emph{Inset})~Level diagram showing competing drive and decay rates
in the system.} \label{Fig:SidebandRabi}
\end{figure}

Sweeping the frequency of a single strong microwave drive on one of
the selective qubit drive lines, and simultaneously measuring the
resonator transmission to carry out a dispersive measurement of the
qubit state \cite{Wallraff2005} shows both the direct qubit
transition at $\omega_{\rm A}$, and an additional spectral line
corresponding to the two-photon blue sideband at $\omega_{\rm
A}^+/2$ (see Fig.~\ref{Fig:SidebandRabi}(a)). The high selectivity
of the individual qubit drive lines suppresses the coupling to qubit
B, the transitions of which are barely visible in
Fig.~\ref{Fig:SidebandRabi}(a). The sideband spectral line is
shifted down at high drive amplitudes due to the AC-Stark effect on
the bare qubit transition \cite{Schuster2005}. At sufficient drive
amplitude, time resolved Rabi oscillations are observed on the blue
sideband transitions (see Fig.~\ref{Fig:SidebandRabi}(b)) by
applying microwave pulses of fixed amplitude and varying length, at
the Stark-shifted two photon blue sideband frequency, to the local
drive line of one qubit. The qubit excited state population is then
measured by applying a microwave pulse to the resonator, and
comparing the time evolution of the transmission to a numerical
solution of the coupled qubit resonator dynamics \cite{Blais2004}.

Fig.~\ref{Fig:SidebandRabi}(b) shows the extracted excited state
population for such an experiment on qubit A (average of
$6.6\times10^4$ repetitions). The solid blue data points show the
results obtained when the system is initially in its ground state
$\ket{g0}$, while the green open data points correspond to an
experiment in which the system is first excited to the state
$\ket{e0}$ using a $\pi$-pulse on the direct qubit transition. In
the latter case, no oscillations are observed since the state
$\ket{e0}$ is not addressed by the blue sideband, a fact that was
also observed in Ref.~\onlinecite{Chiorescu2004}. Also shown in
Fig~\ref{Fig:SidebandRabi}(b) (solid lines) are master equation
simulations of the evolution, using values for the qubit relaxation
rate $\gamma/2\pi=0.2~\rm{MHz}$ and photon decay rate
$\kappa/2\pi=1.7~\rm{MHz}$ taken from separate measurements, and
with the blue sideband transition rate $\Omega/2\pi=8.15~\rm{MHz}$
as a fit parameter.

At long times it can be seen in Fig.~\ref{Fig:SidebandRabi}(b) that
the qubit excited state population tends to a steady state value of
$p_{\rm ss}\approx 0.9$. This can be explained by considering the
different drive and decay channels present in the system (see
inset). Since the state $\ket{e0}$ has no blue sideband transition,
and $\gamma\ll\kappa$, driving the sideband from the ground state
$\ket{g0}$ results in a build up of population in $\ket{e0}$. A
simple model taking into account only the lowest four levels of the
system predicts $p_{\rm
ss}=\kappa\Omega/(\kappa\Omega+\gamma\kappa+\gamma\Omega)=0.87$. It
is worth noting that an increase in the ratio $\kappa/\gamma$ would
allow high fidelity excited state preparation by pumping of the blue
sideband, in a similar manner for example to spin preparation by
optical pumping in semiconductor quantum dots
\cite{Atature2006,Gerardot2008}.

\begin{figure}[t]
\includegraphics[width=1.0 \columnwidth]{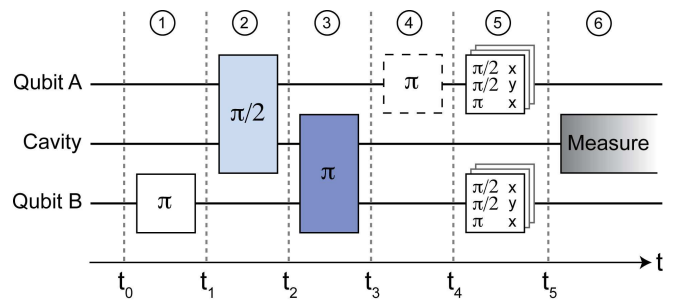}
\caption{Pulse sequence implemented on the two-qubit and resonator
system, to generate and characterize qubit-qubit Bell states. Direct
resonant qubit pulses are shown in white, while blue sideband pulses
between resonator and qubits A and B are shown in light and dark
blue respectively. The ideal state of the system neglecting decay
and imperfections after step $i$, at each of the intermediate times
$t_i$ is indicated in the text.} \label{Fig:PulseSeq}
\end{figure}

Having characterized the  sideband Rabi oscillations for both qubits
A and B, we implement a sequence of pulses on the two-qubit plus
resonator system to generate qubit-qubit entangled states, in a
similar manner to that implemented in trapped ions \cite{Roos2004}.
The full sequence is shown schematically in Fig.~\ref{Fig:PulseSeq}.
At time $t_0$, the system is in its ground state $\ket{gg0}$, and a
resonant $\pi$-pulse is first applied to qubit B, generating the
state $\ket{ge0}$ at $t_1$. A $\pi/2$-pulse on the blue sideband of
qubit A then generates the entangled state
$(\ket{ge0}+e^{i\phi'}\ket{ee1})/\sqrt{2}$. The qubit-resonator
entanglement is then transferred to qubit-qubit entanglement with a
$\pi$-pulse on the blue sideband of qubit B, generating the Bell
state $\ket{\Psi}=(\ket{ge}+e^{i\phi}\ket{eg})/\sqrt{2}$ at $t_3$,
with the resonator returning to its ground state $\ket{0}$. The
phase $\phi$ of the state is dependent on the phase difference
between the two blue sideband pulses in the sequence. An additional
resonant $\pi$-pulse may now be applied to qubit A to generate the
Bell state $\ket{\Phi}=(\ket{gg}+e^{-i\phi}\ket{ee})/\sqrt{2}$ at
$t_4$.

A full tomographic reconstruction of the two-qubit state is carried
out by repeating the pulse sequence 16 times, with different
combinations of additional final single qubit rotations (identity,
$\pi/2_x$, $\pi/2_y$, and $\pi_x$ on each qubit). A microwave pulse
close to the resonator frequency is then used for an averaged
measurement of the two-qubit state. Since the resonator frequency is
shifted by a different amount for each of the four two-qubit states
$\ket{gg}$, $\ket{ge}$, $\ket{eg}$, $\ket{ee}$ \cite{Majer2007}, and
since the resonator transmission quadrature amplitudes depend
non-linearly on this shift, the full density matrix can be
parametrised in terms of the 16 averaged measurement results
\cite{Filipp2008}. The closest physical density matrices to those
experimentally measured are found using a maximum likelihood
approach \cite{Paris2004}.

\begin{figure*}[t]
\includegraphics[width=2.05 \columnwidth]{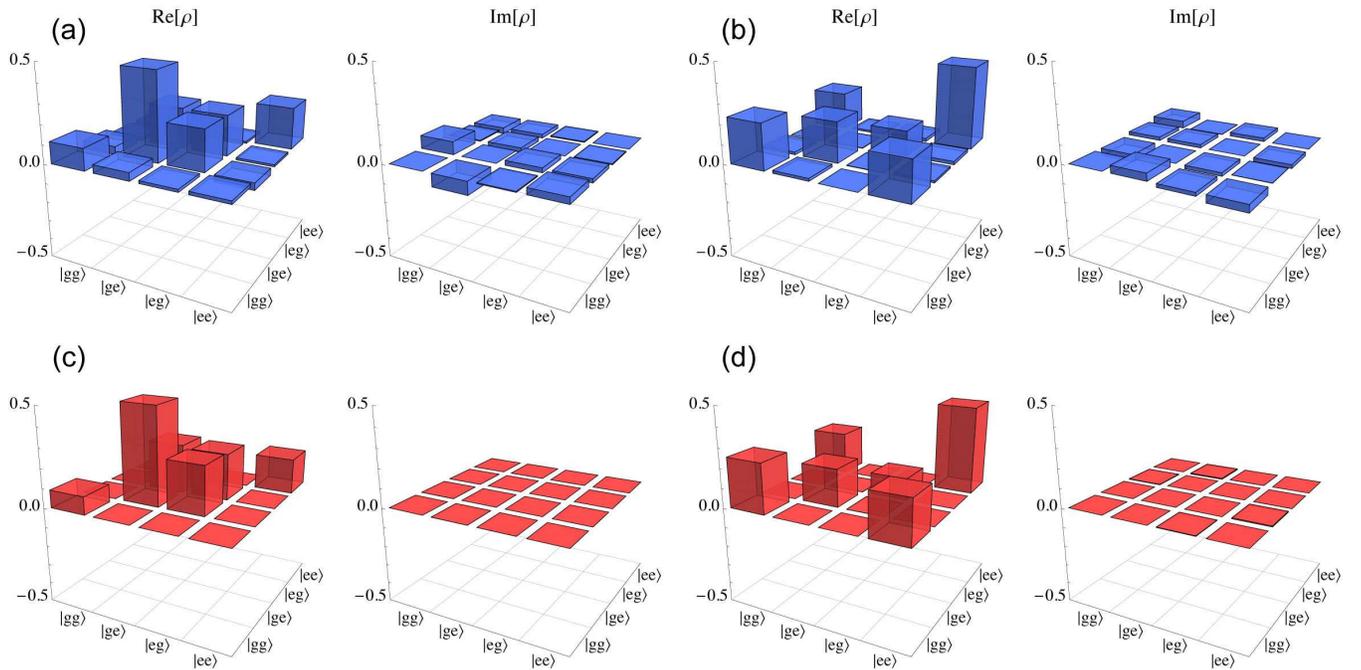}
\caption{(a,b) Real and imaginary parts of the two-qubit density
matrix extracted from two-qubit tomography measurements after
carrying out the pulse sequence to generate the Bell states
$\ket{\Psi_+}$ and $\ket{\Phi_+}$ respectively. (c,d) Simulated
density matrices corresponding to the states shown in (a) and (b)
respectively.} \label{Fig:DMatrices}
\end{figure*}

Examples of experimental two-qubit density matrices $\rho$ measured
after carrying out the pulse sequence for generation of the
$\ket{\Psi_+}=(\ket{ge}+\ket{eg})/\sqrt{2}$ and
$\ket{\Phi_+}=(\ket{gg}+\ket{ee})/\sqrt{2}$ Bell states are shown in
Fig.~\ref{Fig:DMatrices}(a) and (b). In each case $6.6\times 10^5$
repetitions of the experiment were averaged. The fidelities
$\mathcal{F}\equiv(\bra{\psi}{\rho}\ket{\psi})^{1/2}$ of the
measured states $\rho$ with respect to the perfect Bell states $\psi
= \Psi_+$ and $\Phi_+$ are $\mathcal{F} = 74\%$ and $76\%$,
respectively, where $\mathcal{F}$ is limited primarily by the
resonator photon lifetime, $1/\kappa\approx 100~\rm{ns}$.  We have
repeated the state preparation 33 times and calculated the
fidelities separately each time, obtaining an average fidelity of
$75\pm2\%$ for the full set. The entanglement of formation
\cite{Bennett1996} of the generated states extracted from their
density matrices is $0.09 \pm 0.04$.

Also shown in Fig.~\ref{Fig:DMatrices} ((c) and (d)) are the density
matrices resulting from a master equation simulation of the
experiment. The simulation makes both the dispersive and rotating
wave approximations. Two energy levels for each qubit, and the first
5 levels of the harmonic oscillator are taken into account. Energy
relaxation and decoherence rates for the three individual systems
are taken from separate measurements. The fidelity between simulated
and measured density matrices is $97\pm2\%$, indicating that the
loss of fidelity in the experimental state preparation is dominated
by photon loss.

The excellent agreement between experiment and simulation gives
confidence that with a longer photon lifetime, higher fidelity
entangled states are likely to be within reach by this method. The
photon lifetime may be increased in future experiments with the use
of higher quality factor resonators. The scheme may then be
extendable to the implementation of a universal CNOT gate
\cite{Schmidt-Kaler2003,Riebe2006}, and should scale well to a
system with a larger number of qubits. The sidebands may also be
used to generate non-classical states of the microwave field in the
resonator, such as Fock states.

\begin{acknowledgments}
We thank A. Blais, J.~M. Gambetta and H. H\"affner for valuable
discussions and comments on the manuscript. This work was supported
by the Swiss National Science Foundation, by the EC via the EuroSQIP
project and by ETH Zurich. P.J.L. acknowledges support from the EC
via an Intra-European Marie-Curie Fellowship.
\end{acknowledgments}
\bibliographystyle{apsrev}

\end{document}